\documentstyle[prd,tighten,aps]{revtex}

\newcommand{\be}{\begin{equation}}
\newcommand{\ee}{\end{equation}}
\newcommand{\rf}[1]{(\ref{eq:#1})}

\begin{document}
%\draft

\twocolumn[\hsize\textwidth\columnwidth\hsize\csname
@twocolumnfalse\endcsname
%esslett.tex/ \today\\
\title{The accelerating universe in brane-world cosmology}
\author{M. D. Maia\footnote{maia@fis.unb.br}}
\address{Instituto  de F\'\i sica, Universidade de Bras\'\i lia,
 70919-970, Bras\'\i lia, DF, Brazil}
\author{E. M. Monte\footnote{edmundo@fisica.ufpb.br}}
\address{Departamento de F\'\i sica, Universidade Federal da Para\'\i ba,
58059-970, Jo\~ao Pessoa, PB, Brazil}
\author{J. M. F. Maia\footnote{jmaia@axpfep1.if.usp.br}}
\address{IAG, Universidade de S\~ao Paulo, C.P. 3386, 01060-970, S\~ao Paulo,
 SP,
Brazil\\
FINPE, Instituto de F\'\i sica, Universidade de S\~ao Paulo, C. P.
66318, 05315-970, S\~ao Paulo, SP, Brazil}

%\date{\today}

%end  wide text

\maketitle

\begin{abstract}
The standard Friedmann universe embedded in a five dimensional and constant
 curvature bulk
is examined without any a priori junction condition between the brane and the
 bulk.
A geometrical explanation for the accelerated expansion of the universe is
 derived by using
a minimum set of assumptions consistent with the brane-world program. It is
 shown  that the extrinsic
curvature of the brane can be associated to the dark energy which presumably
 drives the universe
expansion.
\end{abstract}
\vspace{3mm}
{PACS number: 04.50+h;98.80cq}

\vskip3mm]

%\section{Introduction}

Recent  observations  of type Ia supernovae \cite{Perlmutter}
suggest that the observable universe is presently undergoing an
accelerated  expansion. Together with other data, this phenomenon  indicates the
 existence of
an energy component characterized by a negative pressure, contributing
 with about 70\% of the total energy density of
the universe (the other 30\% being essentially nonrelativistic matter).
In addition to the most natural candidate for such a component, the cosmological
 constant, many phenomenological  models with a negative pressure, like scalar
 fields with dominant potential energy (quintessence) \cite{5ess} or x-matter
 \cite{xmat}, have been proposed to explain the data.

        Although interesting, in general the dynamics of dark energy candidates
 has no justification but phenomenological.
In particular, as the energy densities of the dark energy and of the other
 components decrease at different
rates and are usually considered as unrelated, the near coincidence in their
 values today can be explained only by either a fine tunning of
initial conditions or a careful choice
 of ``tracking''
potentials \cite{Zlatev}. Alternative proposals
which explain this ``coincidence problem'' consider a phenomenological coupling
 between the dark components,
so that the universe present an asymptotic regime with fixed dark matter to dark
 energy ratio \cite{coupled}. In spite
of any kinematical advantages, it is desirable that a dark energy candidate has
 its origin and particular dynamics
motivated by an underlying theory, in addition to the agreement with the
 observational data.

        Coincidently or not, a new  higher  dimensions gravitational theory
aimed to  solve the hierarchy problem  has gained much attention recently. It
emerged from the observation that the commonly accepted    hypothesis that
gravitation becomes strong  only at around
 $10^{19}$ GeV is a  conjecture
devoid  of   any experimental support, and that   there is  no  theoretical or
experimental  evidence which  prevents   the existence of quantum
gravity effects at the   TeV  scale \cite{ADD}. This so called brane-world
program  is a blend  of some   features of  Kaluza-Klein and  string (or  M)
theories with additional hypotheses.

       In Kaluza-Klein theory all fundamental interactions exhibit   similar behavior  with
respect to  the extra  dimensions  and
this  eventually led to  inconsistencies  in its low energy
 sector. On the other hand, M-theory   based on the Nambu-Goto principle, uses a
 more  flexible  geometrical   condition,
specifying  a classical  space-time  resulting  from    the  motion of an
 extended
object  or 3-brane embedded in the bulk. In one  of its  formulations \cite{HW},
 the standard gauge fields  do not  propagate  along the extra dimensions,  a
 property  which is reserved to the gravitational field.

        In the brane-world program, like in Kaluza-Klein theory,   the geometry of the
higher dimensional  bulk is  a  solution of  the higher dimensional Einstein's
 equations,  but our
space-time (or brane-world) is an embedded  submanifold.  Furthermore,   only
 gravitons  probe the
extra dimensions  at the TeV scale, while the  standard gauge interactions
 remain confined to the
brane \cite{reviews}.

      Several   brane-world cosmologies  have been proposed, mostly  in the
context of the Randall-Sundrum (RS) formulations \cite{RS},
defined in a  five dimensional anti-de Sitter (AdS$_{5}$)  space.  The  dynamics
of these models features     boundary terms  in the
action  and sometimes  mirror symmetry,  such that bulk gravitational waves interfere with the  brane-world motion.  This  usually comes together with   junction conditions   producing an  algebraic  relationship between  the extrinsic curvature and the
confined matter \cite{Carter}. The  consequence is  that
Friedmann's equation acquires  an additional term which is proportional  to the
 square of  energy density of the confined source fields \cite{Binetruy}.
In order to be consistent  with the predictions from  the big-bang
 nucleosynthesis such term  must be non negligible only  at the  very early
 universe, thus having no effect in low energy cosmology.  Some attempts to
 solve this  difficulty have been proposed, using  flat bulks  and appropriate
 boundary conditions \cite{Porrati}. With this,  a late time accelerating
 solution and suitable observational consequences can also be obtained, but at
 the expense of displaying a correcting factor in the Newtonian potential, thus
 introducing phenomenological difficulties \cite{Porrati,Jailson}.

        Such results may give the impression that the  brane-world
 program necessarily depends on  fine tunnings of parameters or on
 the  inclusions of extra fields in order to fit cosmological data.
 The purpose of this note is to show  that even under more general
 conditions, still compatible with the brane-world program, it is
 possible to find a richer set of cosmological solutions in
 accordance to the current observations. In particular, it is found
 that the accelerated expansion of the FRW universe can be
 explained by the dynamics of the extrinsic curvature along the
 radial direction of the brane, thus proposing a geometric
 alternative to the  dark energy problem.

        The  emergence of the squared energy term in Friedmann's
equation is  a consequence of an {\it a priori} specification  of
a junction condition for the discontinuity between the bulk and the
brane. As we shall see later, if such a relation is relaxed, more
general solutions for the integrability conditions of the embedding
equations, in particular the Codazzi's equation, can be obtained.
In order to sort this out, we apply  only  the   relevant features
of a five-dimensional brane-world cosmology in  a  constant
curvature bulk. The analysis is based only on the three  basic
postulates  of the  program, namely, a)the  confinement of standard
gauge interactions, b)the  existence of quantum gravity in the
bulk, and c)the embedding of the brane-world. All other model
dependent properties such  as warped metrics, mirror  symmetries,
radion or extra scalar fields, fine tunning  parameters like the
tension of the brane, and   the  choice of a junction  condition
are left out as much as possible in our calculations.

In the following  we will consider the  FRW  line element
\begin{equation}
ds^{2}=-dt^{2} + a^{2}[dr^{2}+f(r)(d\theta^{2}
 +\sin^{2}\theta d\varphi^{2})]
\end{equation}
where  $f(r)=\sin r, r, \sinh r$   corresponding to   $k=1,0,-1$ respectively.
It is well known  that this space-time can be embedded into a five-dimensional
flat space \cite{Rosen}, but here we extend this embedding to   any  constant
curvature space,   including anti-de Sitter  AdS$_{5}$,  de Sitter dS$_{5}$,
and  the flat M$_{5}$ cases,  with metric  signatures $(4,1)$ and with  the
bulk Riemann tensor\footnote{Greek indices go from 1 to 5 and refer to the
bulk, small  case Latin indices go from 1 to 4 and refer to the brane.}
\begin{equation}
{\cal R}_{\mu\nu\rho\sigma} =  K_{*} ({\cal G}_{\mu\rho}{\cal
 G}_{\nu\sigma}-{\cal G}_{\mu\sigma}{\cal G}_{\nu\rho}) \label{eq:bulk}
\end{equation}
where  ${\cal G}_{\mu\nu}$ is the bulk metric and
  $K_{*}$ denotes the   bulk  constant curvature.  In the flat case
 $K_{*}=0$ and in the
de Sitter and  anti-de Sitter cases we may   write  $K_{*}=
 \pm\frac{\Lambda_{*}}{6}$ respectively,
where  $\Lambda_{*}$  is  the bulk cosmological constant.

   In the  normal Gaussian frame defined by the embedded space-time,  the  bulk
 metric may be  decomposed as
\be
 {\cal G }_{\mu\nu}=
\left(\matrix{
g_{ij}  & 0\cr
0 & g_{55}
}\right),\;\;\;  g_{55}=1
\ee
Inserting this in  \rf{bulk} and using the  Gauss and Codazzi
equations, which are the integrability conditions for the embedding
equations \cite{Eisenhart}, we  obtain
\begin{eqnarray}
R_{ijkl} &=&  (k_{ik}k_{jl}- k_{il}k_{kj}) +  K_{*}
 (g_{ik}g_{jl}-g_{il}g_{kj})\label{eq:G1}\\
k_{ij;k} & = & k_{ik;j}
\label{eq:C1}
\end{eqnarray}
where  $k_{ij}$  denotes the extrinsic curvature.

        From the contractions of (\ref{eq:G1})  with  $g^{ij}$ we
obtain the Ricci scalar $R$ of the brane metric, in terms of the
Ricci scalar of the bulk ${\cal R}$. Therefore,  the embedding and
the   supposition that the bulk is  a solution of Einstein's
equations necessarily introduce  an  Einstein-Hilbert  component
in the  brane-world Lagrangian.  After  adding  a  confined matter
Lagrangian  and a   four-dimensional  cosmological constant
$\Lambda$, the  effective Lagrangian compatible with the embedding
reads (for details, see \cite{ME}):
\begin{eqnarray}
{\cal L}_{\rm eff} = &R &\sqrt{-g}    +(K^{2}  +  h^{2})\sqrt{-g}\nonumber\\ &+&{\cal R}\sqrt{-g} -2\frac{d
 h}{d y}\sqrt{-g} +\Lambda\sqrt{-g} -{\cal L}_{m}
\end{eqnarray}
where  $h=g^{ij}k_{ij}$ is the mean curvature  of the brane-world,
$K^{2}=k^{ij}k_{ij}$,   $y$ denotes the fifth  coordinate  and
${\cal R}=-20K_{*}$  as  derived   from \rf{G1}.  The total
derivative  term with respect to  $y$ can be eliminated  provided
the motion of the  brane-world    occurs between  two  fixed
minimal boundary  hypersurfaces where  $h=0$ \cite{ME}.  Like in
the  second RS model such boundaries  can be moved away so that all
boundary  generated bulk waves  are eliminated. Variation of the
action with respect to  $g_{ij}$  gives the dynamical  equation
compatible  with the embedding  and  with the  confined  matter
hypotheses.  Denoting by $T^{m}_{ij}$ the confined matter
energy-momentum tensor and defining $\lambda = -3K_{*}+\Lambda$,
the resulting  Einstein's equations  are
\begin{equation}
R_{ij}\! -\!\frac{1}{2}Rg_{ij} +\lambda g_{ij}  =-\! 8\pi G T_{ij}^{m}\!+
 Q_{ij}\label{eq:EinsteinFRW}
\end{equation}
where
\begin{equation}
 Q_{ij}  = g^{mn}k_{im}k_{jn}-hk_{ij}  -\frac{1}{2}(K^{2}-h^{2})g_{ij}
\label{eq:Qij}
\end{equation}
Notice that this quantity is identically conserved  in the sense that,
\be\label{eq:conserva}
Q^{ij}{}_{;j} = 0,
\ee
so that there is no exchange of energy between this geometrical correction and
the confined matter. Such an aspect has one important consequence: if $Q_{ij}$
is to be related to the dark energy, as we do later, it does not exchange energy
with the ordinary matter, like in the coupled quintessence  models
\cite{coupled}.

 In order to specify a cosmological model, it is usual to
add a  condition on the extrinsic  curvature, such as the  Israel-Lanczos
junction conditions, thus introducing a  relationship  between  $k_{ij}$ and the
confined  energy-momentum tensor \cite{Carter}. Here we have removed or at the
least postponed all motivations, such as  boundary bulk waves and  mirror symmetries,  which
justify the application of such conditions (see, for example, Refs.
\cite{Binetruy}). Alternatively, we solve   Codazzi's  equation  for
any homogeneous and isotropic brane before applying  further constraints on
$k_{ij}$.  Although such a solution does not provide any explicit
cosmological model, it clarifies the role played by the junction condition and
it has  a  degree of freedom  which permits  an adjustment with  the
observational results.

        In order to solve Codazzi's equation, we first notice that for the FRW metric
 $k_{ij}$ is diagonal. Denoting the  spatial indices in the brane by the letters
 $a, b, c, d = 1\ldots 3$, we find that Eq. (\ref{eq:C1}) is reduced to
 \cite{ME}
\begin{eqnarray*}
 k_{aa,c}&- & k_{ad}\Gamma^{d}_{ac}= k_{ac,a}
-k_{cd}\Gamma^{d}_{aa}, \\
{k}_{aa,4}&- &k_{aa}\frac{\dot a}{a} =-a\dot{a}(\delta^{1}_{a}\delta^{1}_{b}
 +f^{2}\delta^{2}_{a}\delta^{2}_{b} +f^{2}\sin^{2}\theta \;
 \delta^{3}_{a}\delta^{3}_{b})k_{44}.
\end{eqnarray*}
The first equation  for  $c\neq 1$  gives  $k_{11,c} =0$ so that
$k_{11}$ is a function $b(t)$. From the second  equation  we obtain
$k_{44}= \frac{-1}{\dot{a}}  \frac{d}{dt}(b(t)/a)$. Repeating the same  arguments
for  $k_{22}$ and  $k_{33}$ we  obtain the general solution of  \rf{C1} \cite{MS}
\begin{equation}
k_{ab}=\frac{b}{a^{2}}g_{ab},\; \mbox{and} \;\; \;\; k_{44}=
 -\frac{1}{\dot{a}}\frac{d}{dt}\left(\frac{b}{a}\right)\label{eq:kij}.
\end{equation}
Denoting  $B= \dot{b}/b$ and  as usual  $H=\dot{a}/a$  we obtain
\begin{equation}
Q_{ab}\!=\! \frac{b^{2}}{a^{4}}\left(2\frac{B}{H}-1 \right)g_{ab},\;
 \;Q_{44}\!=\! -\frac{3b^{2}}{a^{4}}. \label{eq:Qab}
\end{equation}
Replacing (\ref{eq:Qab}) in  \rf{EinsteinFRW}   we obtain   Friedmann's
 equation  modified by the  presence of the extrinsic curvature
\begin{equation}
\dot a^2+k=\frac{8\pi G}{3}\rho  a^{2}+\frac{\lambda}{3}a^{2}+ \frac{b^2}{a^2}
 \label{eq:Friedmann}
\end{equation}

Notice that  in the context of brane-worlds  the geometry of  the
universe  must  show  quantum fluctuations. This  will provide the
necessary   equation   to determine  $b(t)$. While  a  quantum theory
of brane-world  fluctuations is  not available, we may  use classical
perturbative   approaches  for  the generation of embedded geometries
to generate   models  based on the  extrinsic  curvature \cite{ME}.

Just for the sake of  comparision  with  existing models,  consider that
this classical  relation is  replaced by  the Israel-Lanczos  condition
applied to our  solution  \rf{kij},
\be
k_{ij}  = -\frac{1}{2}\alpha_{5}^{2}(T^{m}_{ij}
-\frac{1}{3}T^{m}g_{ij}),
\ee
where $\alpha_5$ is proportional to the gravitational constant in
the bulk. In such a case, we obtain
$b(t)=-\frac{1}{6}\alpha_{5}^{2}
\gamma\rho a^{2} $. By replacing  such an expression in \rf{Friedmann}  it follows that
\be
\dot a^2+k =\frac{8\pi G}{3}\rho  a^{2}+{\alpha_{5}^{2}\gamma^{2} \over 36} \rho^{2} a^{2},
\ee
showing  that  the a $\rho^{2}$ term emerges from the  imposition
of the junction condition. As  already mentioned, this particular
case either does not agree with the observations or require extra
parameters and fine tunnings, consequently we  will will not
consider it in our analysis.

 We  see that  our  solution  \rf{kij} depends  only on the radial
bending  function $b(t)$ which remains arbitrary.  To find the
dynamical role of this function, we associate  $Q_{ij}$ to a
separatedly conserved energy-momentum (from \rf{conserva})
\be
\tau_{ij} \equiv -\frac{Q_{ij}}{8\pi G}.
\ee
Denoting by  $p_{b}$ the  field  pressure,  the bending tension (if negative),
and  by  $\rho_{b}$ the  corresponding bending energy,  we can represent
$\tau_{ij}$  as
\be
\tau_{ij} =(p_{b} + \rho_{b}) U_{i}U_{j} +p_{b}g_{ij}, \;\;
 U_{i}=\delta^{4}_{i}
\ee
In addition, we can also  write a  state-like  equation $p_{b}=
(\gamma_{b} -1)\rho_{b}$  where   $\gamma_{b}$  is an undetermined
function of time. Comparing    $\tau_{ab}$ and $\tau_{44}$  with
\rf{Qab}, we obtain
\be %\begin{eqnarray}
\rho_{b}= \frac{3}{8\pi G}\frac{b^{2}}{a^{4}}, \;\;  p_{b}  = - \frac{1}{8\pi
 G}\frac{b^{2}}{a^{4}}\left(  2\frac{B}{H}-1\right) \label{eq:rho}
\ee %\end{eqnarray}

To determine  the   equation for  $b(t)$,    we use the   trace
$Q=g^{ij}Q_{ij}=(6b^{2}/a^{4})(B/H)$. The   spatial trace is
\be
 g^{ab}Q_{ab} = Q + Q_{44}= 3\frac{b^{2}}{a^{4}}
\left(2\frac{B}{H}-1\right)
\ee
On the other hand, from  the  expression  of  $\tau_{ij}$  we find
\be
g^{ab}Q_{ab} = - 24\pi G p_{b}
\ee
Comparing the  last two expressions  we obtain the   equation
\be
\frac{\dot{b}}{b}= \frac{1}{2}(4-3\gamma_{b})\frac{\dot{a}}{a}
\ee
Notice that this resembles one of the phenomenological candidates
for dark energy, the x-matter \cite{xmat}, but in our case this
field has a fundamental (geometrical) justification for the
equation of state.

        As a simple  example  consider  the case where $\gamma_{b}$
 is  constant. In such a case the above equation yields to  a very
 simple solution
\be
b(t) =b_{0}  a(t)^{\frac{1}{2}(4-3\gamma_{b})}
\ee
where  $b_{0}$ is an integration constant.
With this  solution, the bending energy becomes
\be
\rho_{b} = \frac{3b_0}{8\pi G}a^{-3\gamma_{b}}
\ee
For a rough estimate consider a vanishing $\lambda$ (as, for
example, for  a M$_5$ bulk with vanishing $\Lambda$) and a
spatially flat ($k=0$) brane composed mainly by dark matter and the
bending contribution as the  dark energy. In such a case, the
deceleration parameter reads
\be
q =-\frac{\ddot{a}a }{\dot a^2}= (3\gamma_{b} - 2){\Omega_{b} \over 2} + \frac{\Omega_{m}}{2},
\ee
where $\Omega_i \equiv 8\pi G \rho_i /3H^2$. For $\Omega_m \sim
0.3$ and $\Omega_b \sim 0.7$, as suggested by recent observations,
a present time universe driven by  the bending occurs whenever
$\gamma_{b} < 0.52$ as in the x-matter case \cite{xmat}. In fact,
any observational test suited to test x-matter models can be used
for this ansatz with a constant $\gamma_b$. For the general case,
reconstructing techniques \cite{Saini} can be used in order to find
the appropriate evolution law for $b(t)$. Alternatively,
$\gamma_b(t)$ can be provided by fundamental physics arguments.

        In addition to the dark energy problem, other important
issues can be addressed in our geometrical approach. For example,
primordial inflation can be more natural in this context than in
the usual brane-world proposal with a squared energy density.
Additionally, possible relations between metric fluctuations  and
the extrinsic curvature of the brane can be obtained directly from
geometric constraints and the specific models found can be
considered to study the evolution of cosmological perturbations by
adapting existing formalisms (see for example \cite{fluctuations}).
Other theoretical alternatives, including different bulk signatures
as well as a comprehensive discussion on the topics covered here
will be presented  elsewhere.

        To summarize, we have  shown that the extrinsic curvature
may be associated to the unknown dark energy component that drives the
 acceleration of the universe. The solutions obtained permit a
richer set of phenomenological possibilities. Once available, the
newest observational data can be used to reconstruct the equation
of state of the dark component, thus providing the time evolution
of the brane bending. In contrast with the usual dark energy
models, the decoupling between the dark components is derived,
instead of assumed. The geometrical approach considered here allows
the inclusion of any constant curvature bulk with the same end
result except for a fine tunning between the cosmological constants
of the bulk and of the brane. Our  overall conclusion is that the
dark energy paradigm can be improved by considering a non trivial
contribution of geometrical origin, if a junction condition is not
postulated {\em a priori}.

\end{document}